\documentclass[]{emulateapj}

\usepackage{epsfig} 

\shorttitle{Clouds on GJ1214b}
\shortauthors{Charnay et al.}

\begin{document}


\title{3D modeling of GJ1214b's atmosphere:  \\formation of inhomogeneous high clouds and observational implications}


\author{B. Charnay\altaffilmark{1,2,3}, V. Meadows\altaffilmark{1,2},  A. Misra\altaffilmark{1,2}, J. Leconte\altaffilmark{4,5} and G. Arney\altaffilmark{1,2}}

\affil{\altaffilmark{1}{Astronomy Department, University of Washington,
    Seattle, WA 98125, USA}\\
\altaffilmark{2}{NASA Astrobiology Institute’s Virtual Planetary Laboratory, Seattle, WA 98125, USA}\\
\altaffilmark{3}{NASA Postdoctoral Program Fellow}\\
\altaffilmark{4}{Canadian Institute for Theoretical Astrophysics, University of Toronto, Toronto, ON M5S3H8, Canada.}\\
\altaffilmark{5}{Center for Planetary Sciences, Department of Physical and Environmental Sciences, \\
University of Toronto Scarborough, Toronto, ON M1C 1A4, Canada.}  }

\email{bcharnay@uw.edu}




\begin{abstract}

The warm sub-Neptune GJ1214b has a featureless transit spectrum which may be due to the presence of high and thick clouds or haze. Here, we simulate the atmosphere of GJ1214b with a 3D General Circulation Model for cloudy hydrogen-dominated atmospheres, including cloud radiative effects.  We show that the atmospheric circulation is strong enough to transport micrometric cloud particles to the upper atmosphere and generally leads to a minimum of cloud at the equator. By scattering stellar light, clouds increase the planetary albedo to 0.4-0.6 and cool the atmosphere below 1 mbar. However, the heating by ZnS clouds leads to the formation of a stratospheric thermal inversion above 10 mbar, with temperatures potentially high enough on the dayside to evaporate KCl clouds. We show that flat transit spectra consistent with HST observations are possible if cloud particle radii are around 0.5 $\mu$m, and that such clouds should be optically thin at wavelengths $>$ 3 $\mu$m.  Using simulated cloudy atmospheres that fit the observed spectra we generate transit, emission and reflection spectra and phase curves for GJ1214b.   We show that a stratospheric thermal inversion would be readily accessible in near and mid-infrared atmospheric spectral windows. We find that the amplitude of the thermal phase curves is strongly dependent on metallicity, but only slightly impacted by clouds.  Our results suggest that primary and secondary eclipses and phase curves observed by the James Webb Space Telescope in the near to mid-infrared should provide strong constraints on the nature of GJ1214b's atmosphere and clouds.

\end{abstract}



\section{Introduction}

As a fundamental planetary phenomenon, clouds are expected to play a major role in the chemistry, thermal structure and observational spectra of extrasolar planets \citep{marley13}. 
Clouds have clearly been detected on hot Jupiters \citep{pont13, demory13}.
But they are likely more common on smaller planets containing a larger fraction of heavy elements and condensable species in their atmospheres. Featureless transmission spectra suggesting the presence of clouds or haze have been obtained for one warm Neptune \citep{knutson14a} and two warm sub-Neptunes \citep{kreidberg14a, knutson14b}. 
In particular, GJ1214b is a warm mini-Neptune or waterworld (i.e. with a hydrogen-rich or a water-rich atmosphere) orbiting around a nearby M dwarf. Measurements by the Hubble Space Telescope (HST) revealed a very flat spectrum between 1.15 and 1.65 $\mu$m \citep{kreidberg14a}. This has been interpreted as the presence of high clouds or haze with a cloud-top at around 0.1-0.01 mbar, depending on the atmospheric metallicity \citep{kreidberg14a}.
A Titan-like organic haze may be photochemically produced, but the actual mechanism is not yet known for such a warm atmosphere.
Condensate clouds of potassium chloride or zinc sulfide (KCl and ZnS) may also form \citep{miller-ricci12, morley13}, but would form deeper in the atmosphere, at 0.1-1 bar (see Fig. \ref{figure_1}a). A featureless transit spectrum due to condensate clouds would therefore require a strong atmospheric circulation, lofting particles to lower pressures. 

Here we use the observed spectra of GJ1214b as a test case to constrain mechanisms for the possible formation of high condensate clouds on GJ1214b and similar exoplanets. We use a three-dimensional General Circulation Model (GCM), simulating GJ1214b for H-dominated atmospheres with KCl and ZnS clouds. In section \ref{Model}, we describe the model. In section \ref{Results}, we present our results concerning the cloud distribution and the atmospheric thermal structure. In section \ref{Observational spectra}, we describe the impact of clouds on the spectra. Finally, we discuss all these results and their implications for future observations in section \ref{Discussion}.

\section{Model} \label{Model}

\subsection{The Generic LMDZ GCM} 
We performed simulations of GJ1214b's atmosphere using the Generic LMDZ GCM \citep{wordsworth11, charnay15a}. The model and simulations for cloud-free atmospheres of GJ1214b are fully described in \cite{charnay15a}. The radiative transfer is based on the correlated-k method. k-coefficients are computed using high resolution spectra from HITRAN 2012 \citep{rothman13} and HITEMP 2010 \citep{rothman10}.
Here, we considered a H$_2$-rich atmosphere at 100$\times$solar metallicity, which we found optimal for the cloud vertical mixing \citep{charnay15a}. The composition is assumed to be at thermochemical equilibrium in each GCM cell. We used the same orbital, physical and stellar parameters as in \cite{charnay15a}. In particular, we assumed that GJ1214b is synchronously rotating.

We ran simulations with a 64$\times$48 horizontal resolution and with 50 layers equally spaced 
in log pressure, spanning 80 bars to 6$\times$10$^{-6}$ bar.
Simulations were started from a 1D temperature profile computed with the 1D version of the model. We ran simulations without cloud radiative effects for 1600 days. The simulations with cloud radiative effects required a radiative timestep 10 times shorter. In these cases, we ran simulations for only 300 days. To accelerate the convergence, the first 200 days were run by increasing the radiative heating/cooling by a factor proportional to the pressure when it is higher than 0.1 bar. This technique resulted in simulations close to equilibrium after 300 days, with a relative difference between total emitted radiation and total absorbed radiation less than 1$\%$.

\subsection{Cloud microphysics and optical properties} 
We considered only KCl and ZnS clouds, although Na$_2$S and other iron or silicate clouds could form in the deeper atmosphere and be removed \citep{morley13}.
We fixed the radius of cloud particles everywhere in the atmosphere, and treated the cloud particle radius as a free parameter. We ran simulations with cloud particle radii from 0.1 to 10 $\mu$m, with the same radii for KCl and ZnS clouds.
The abundance of cloud particles evolves through condensation, evaporation, transport and sedimentation. 

The formation of KCl and ZnS cloud occurs through the thermochemical reactions \citep{morley12}:
\begin{equation} \label{equation1}
\rm KCl = KCl_{(s)}
\end{equation} 
\begin{equation} \label{equation2}
\rm H_2S +Zn = ZnS_{(s)}+H_2
\end{equation}

We used the saturation vapor pressures of KCl and ZnS from \cite{visscher06} and \cite{morley12}.
For simplicity, we considered only the transport of atomic Zn (the limiting element in equation 2) as an idealized ZnS vapor.  We then considered both reactions as simple gas-solid phase changes with no supersaturation and with a latent heat of  2923 kJ/kg for KCl and 3118 kJ/kg for ZnS.
We fixed the abundance of KCl vapor to 2.55$\times$10$^{-5}$ mol/mol (4.3$\times$10$^{-4}$kg/kg) and the abundance of ZnS vapor to 8.5$\times$10$^{-6}$ mol/mol (1.8$\times$10$^{-4}$kg/kg) in the deep atmosphere, corresponding to values from \cite{lodders03} for a 100$\times$solar metallicity.  Cloud particles were assumed to be spherical, with a density of 2000 kg/m$^3$ (KCl) and 4000 kg/m$^3$ (ZnS), and to sediment at a terminal velocity given in \cite{charnay15a}.

We computed the optical cloud properties for the different radii using a log-normal size distribution and optical indices from \cite{querry87}. KCl is purely scattering in the visible and near-infrared while ZnS has large absorption bands centered near 0.2 and 1 $\mu$m.

\section{Results} \label{Results}

\begin{figure}[!h] 
\begin{center} 
	\includegraphics[width=8.9cm]{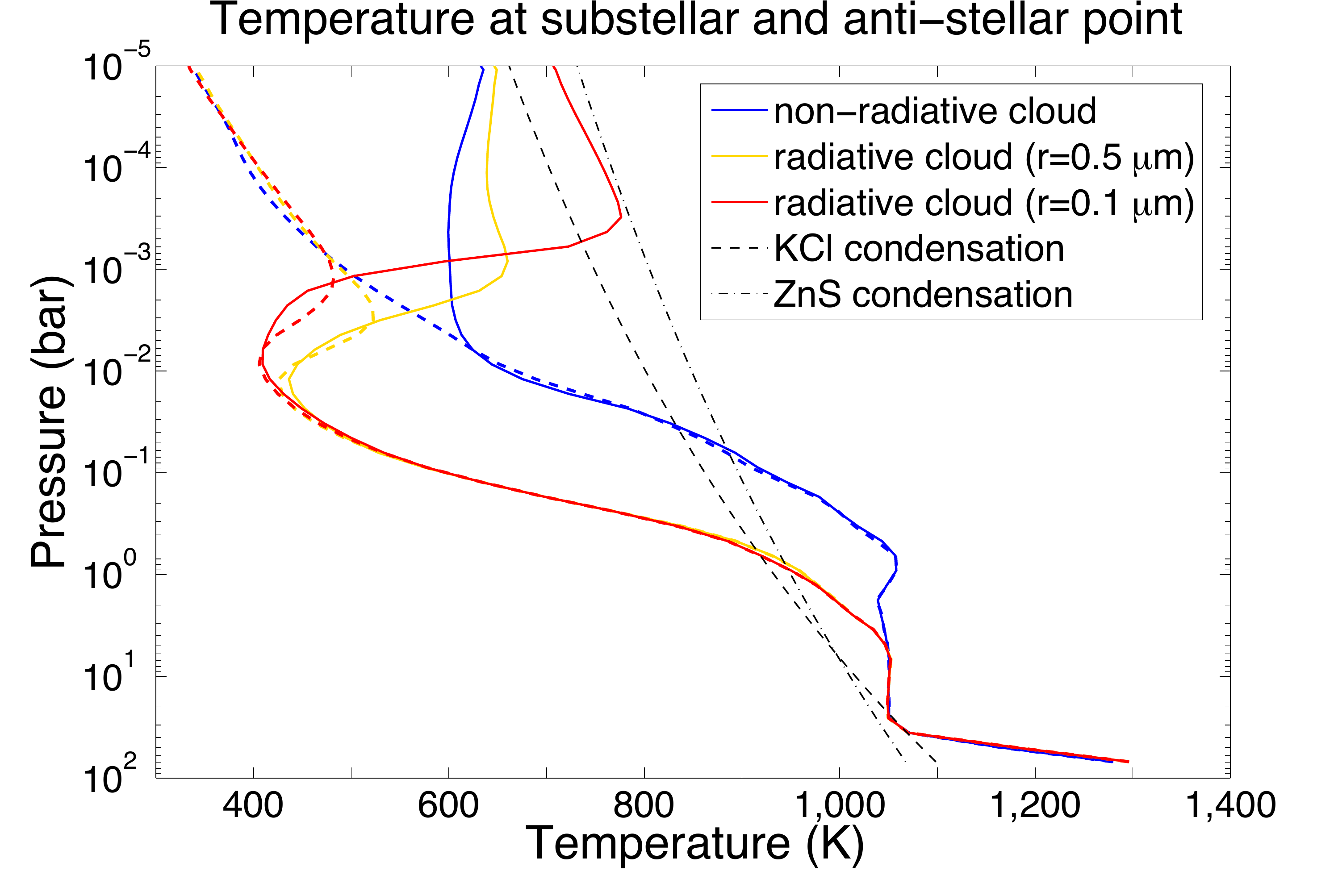}
	\includegraphics[width=8.9cm]{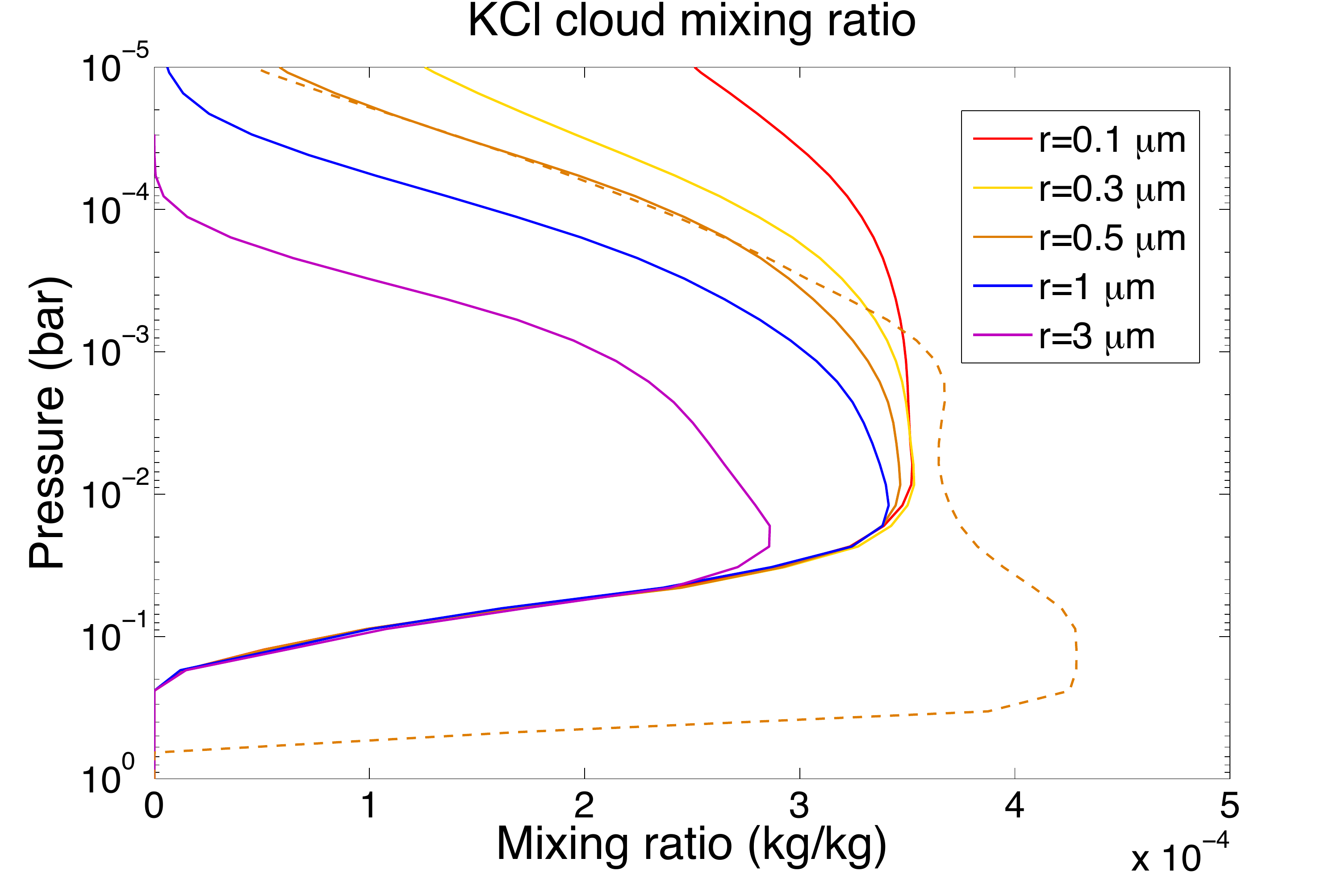}
\end{center} 
\caption{Top panel: temperature profiles at the substellar point (solid color lines) and the anti-stellar point (dashed color lines) for the 100$\times$solar metallicity composition with non-radiative cloud (blue) and with radiative cloud particles with radii of 1 $\mu$m (orange) and 0.1 $\mu$m (red). Dashed and dashed-dotted black lines correspond to the condensation curves of KCl and ZnS.
Bottom panel: KCl cloud mixing ratio for particle radii from 0.1 to 3 $\mu$m. The dashed line corresponds to the case with radiatively active cloud.} 
\label{figure_1}
\end{figure} 

\subsection{Simulations with non-radiatively active clouds} 
\begin{figure*}[t] 
\begin{center} 
	\includegraphics[width=8.5cm]{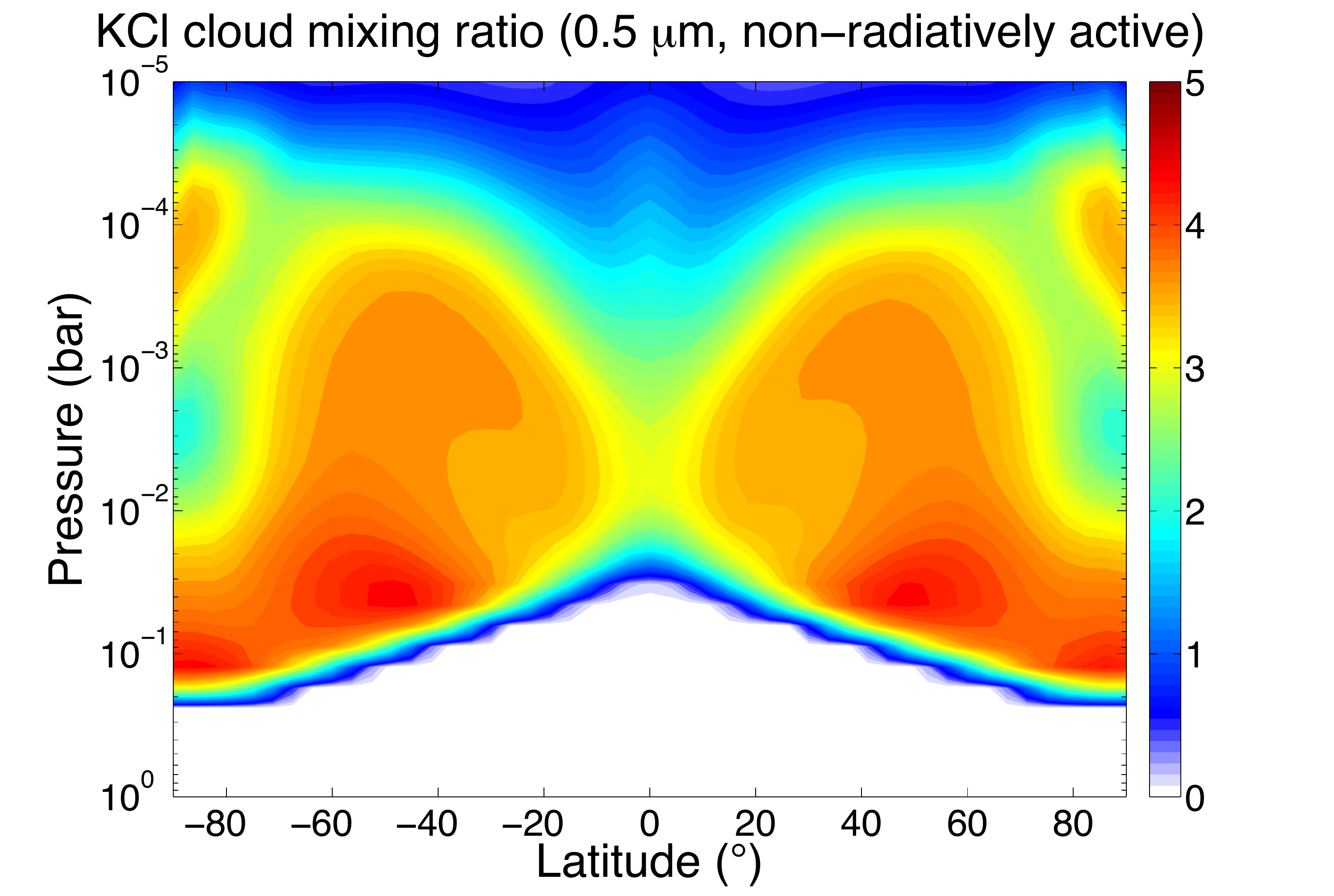}
	\includegraphics[width=8.3cm]{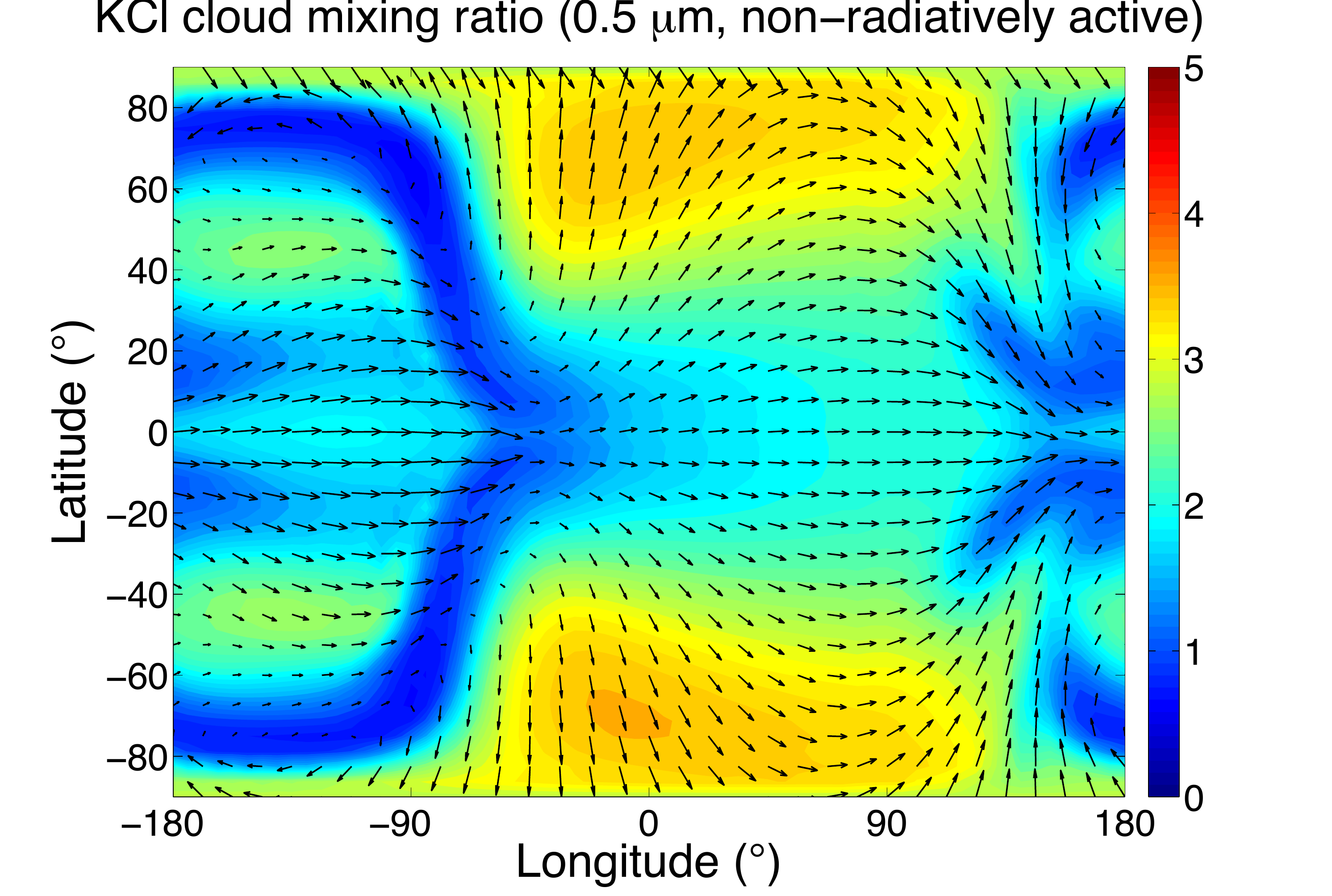}
	\includegraphics[width=8.5cm]{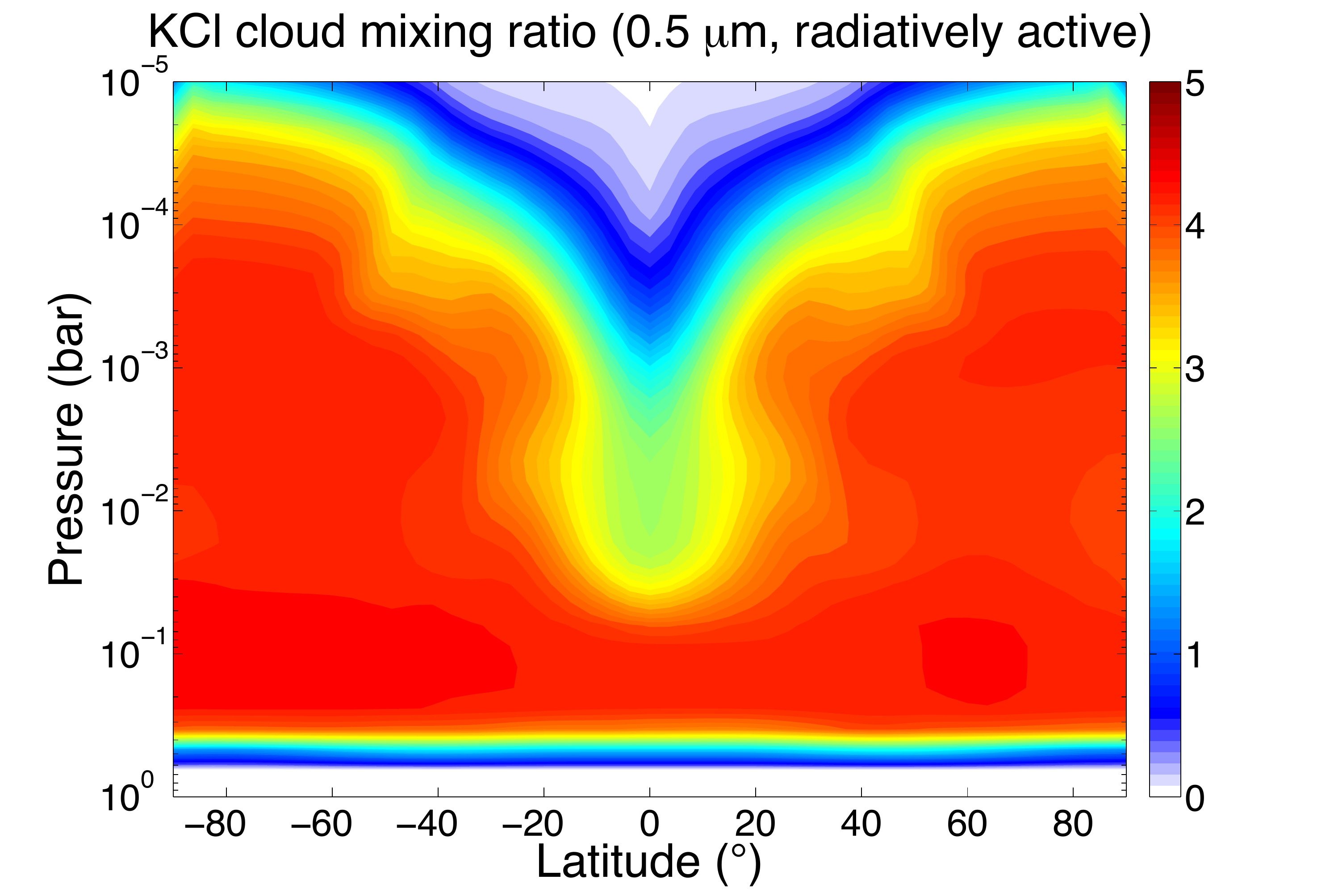}
	\includegraphics[width=8.3cm]{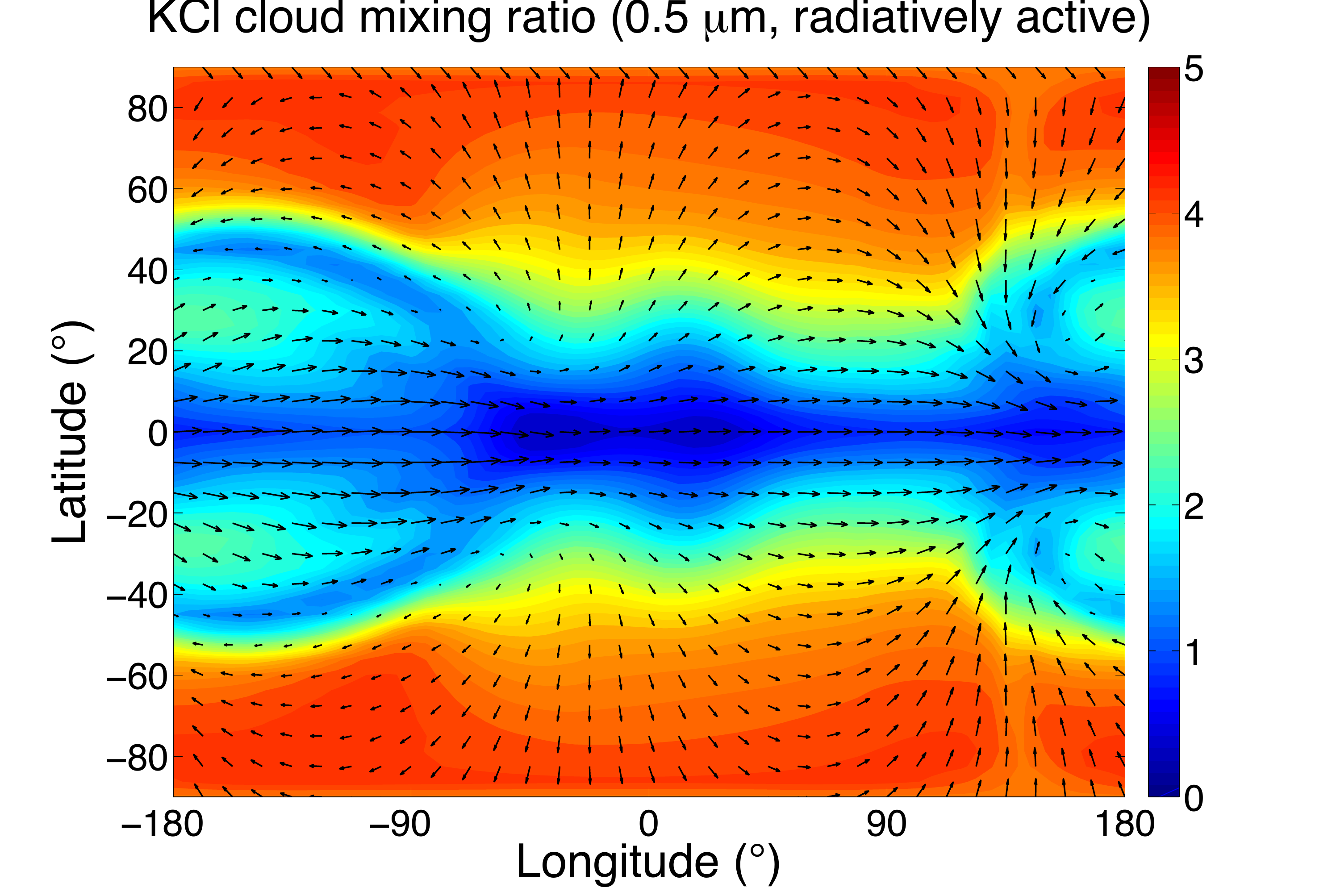}
	\includegraphics[width=8.5cm]{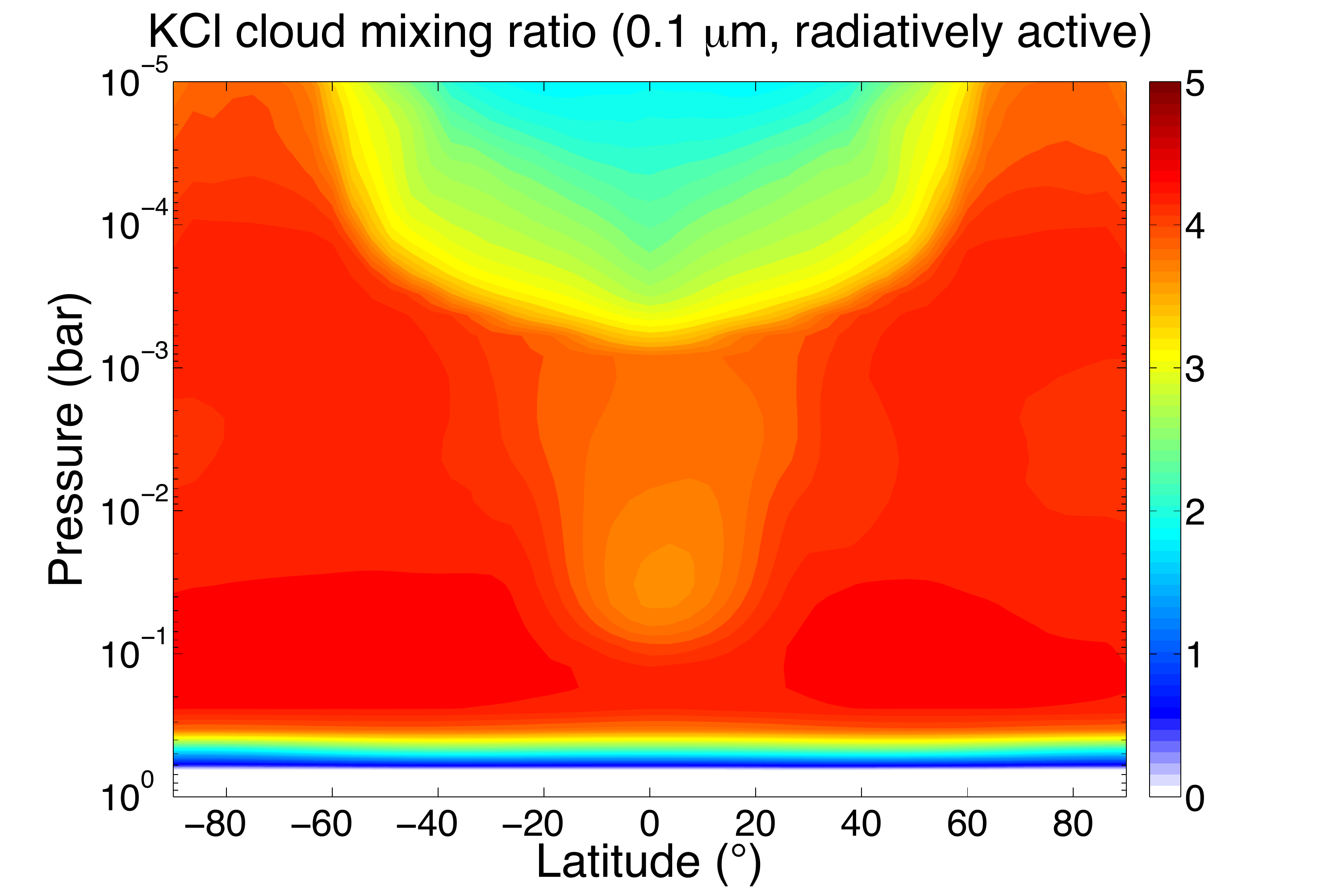}
	\includegraphics[width=8.3cm]{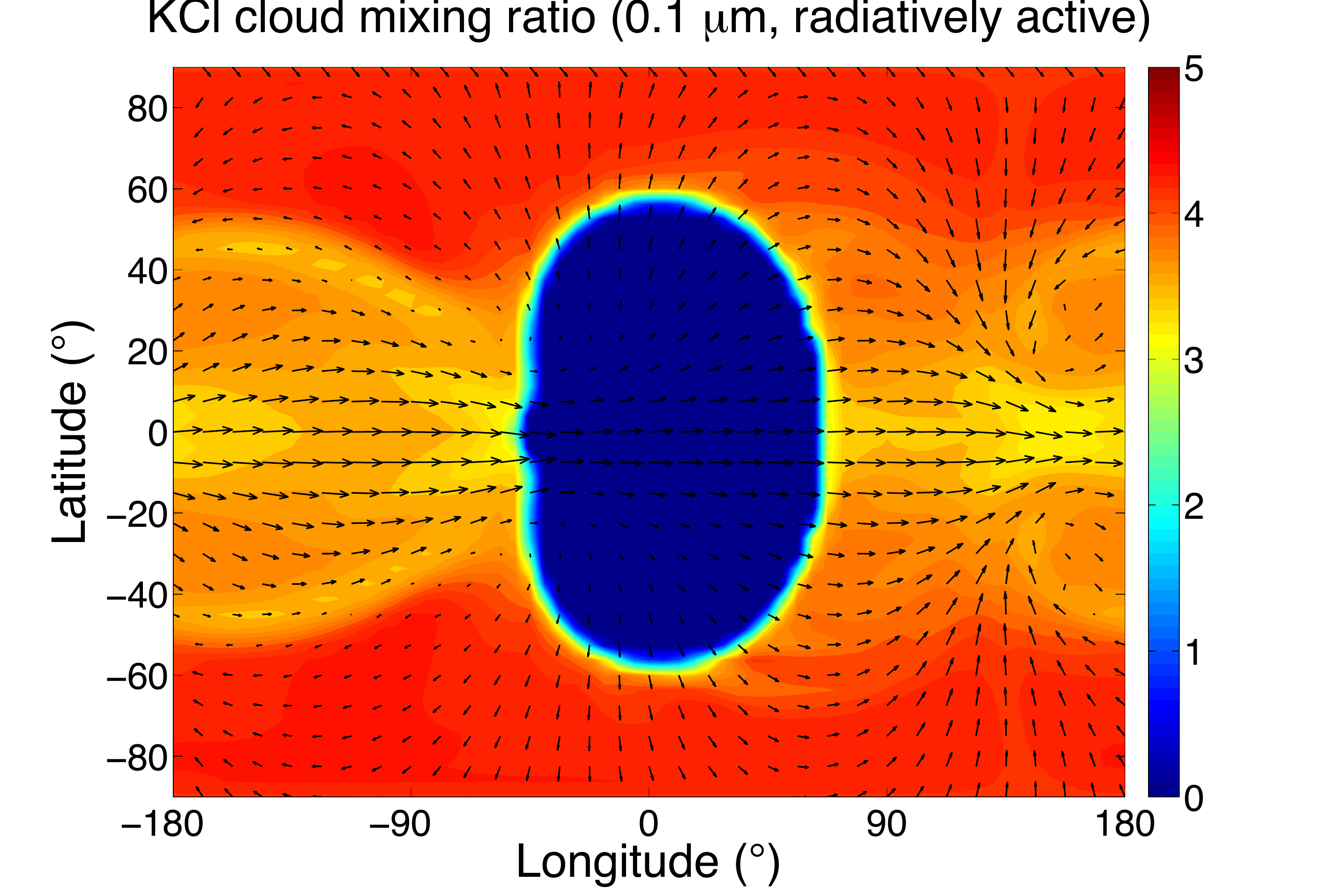}
\end{center}  
\caption{Zonally averaged KCl cloud mixing ratio (left panels) and  KCl cloud mixing ratio at 0.1 mbar (right panels) in 10$^{-4}$kg/kg. 
Top panels are for non-radiatively active clouds with 0.5 $\mu$m particles.
Middle panels are for radiatively active clouds with 0.5 $\mu$m particles.
Bottom panels are for radiatively active clouds with 0.1 $\mu$m particles.}
\label{figure_2}
\end{figure*}

We first ran simulations with non-radiatively active clouds. The latent heat release due to cloud condensation has no impact on the atmospheric dynamics, as is seen in the cloud-free case described in \cite{charnay15a}. Temperature variations between the dayside and the nightside mostly occur above 10 mbar (see Fig. \ref{figure_1}a). The altitude above which clouds form drops from around 40 mbar at the equator to around 200 mbar at the poles because of latitudinal temperature gradients (see Fig. \ref{figure_2}). Above this condensation altitude, temperatures are never high enough to significantly evaporate clouds. Cloud particles are therefore stable and act as simple tracers. Because the sedimentation timescale is generally much longer than the advection timescale, their distribution is mostly driven by the zonal mean circulation. For a synchronously-rotating planet, the circulation is shaped by the strong day-night temperature contrast and standing planetary waves \citep{showman11}. This pattern is globally associated with upwelling on the dayside and equatorial downwelling on the nightside \citep{charnay15a}. The latter is mostly produced by two very strong equatorial downdrafts close to the terminators and dominates below 1 mbar for 100$\times$solar metallicity. Therefore, the zonal mean circulation globally exhibits an anti-Hadley circulation below 1 mbar and a Hadley circulation above \citep{charnay15a}.
This anti-Hadley circulation occurs in the region where clouds form, and strongly impacts their global distribution, leading to a cloud minimum at the equator and a maximum between 40-60$^\circ$ latitude in both hemispheres for pressures lower than 0.1 mbar (see top left panel in Fig \ref{figure_2}).
This equatorial minimum is reinforced above 1 mbar by a net poleward cloud transport (see top right panel in Fig \ref{figure_2}).
Fig. \ref{figure_1}b shows the global mean mixing ratio of condensed KCl cloud for particle radii from 0.1 to 3 $\mu$m. Cloud particles are well mixed to high altitudes for radii of 0.1 $\mu$m. For particles larger than 1 $\mu$m, the upper atmosphere (above 10 mbar) is depleted in cloud. 

\subsection{Simulations with radiatively active clouds}
The 3D modeling of radiatively active clouds in GJ1214b's atmosphere is challenging. Because of cloud opacity, the atmospheric radiative timescale becomes shorter in the upper atmosphere and the condensation occurs deeper in the atmosphere, where the sedimentation timescale is longer. 
With the current computational resources, it is impossible to obtain a fully converged 3D simulation at pressure significantly greater than around 1 bar. However, all results (e.g. cloud vertical mixing, thermal structure and observational spectra) for the upper atmosphere can be valid.
We ran simulations with radiatively active clouds for 300 days, accelerating thermal convergence (see section \ref{Model}). These simulations were thermally converged up to around 1 bar and the cloud distribution is locally at equilibrium up to 10 mbar (see \cite{charnay15a}). Since the simulations were not converged in the deep atmosphere, they may overestimate cloud abundance in the upper atmosphere.

With radiatively active clouds, the planetary albedo increases from almost 0 to 0.4 for 3 $\mu$m particles, and up to 0.6 for 0.1 $\mu$m particles. Stellar radiation penetrates less deeply in the atmosphere, cooling pressures higher than $\sim$1 mbar (see Fig. \ref{figure_1}a). Consequently, clouds form deeper, at around 0.6 bar for 0.5 $\mu$m particles. At this pressure, there is almost no latitudinal temperature variation and so clouds form at the same altitude globally (see Fig. \ref{figure_2}). Above 10 mbar, the cloud vertical mixing is similar with and without cloud radiative effects (see Fig. \ref{figure_1}b). With radiatively active clouds, the anti-Hadley circulation appears reinforced with a clear minimum of cloud at the equator (see Fig. \ref{figure_2}).

Above 10 mbar, the absorption of stellar radiation by ZnS clouds forms a stratospheric thermal inversion on the dayside. At the substellar point, the difference between the maximal and the minimal stratospheric temperature reaches around 350 K, 225 K and 50 K for particle radii of 0.1, 0.5 and 1 $\mu$m respectively. In contrast, the temperature at the anti-stellar point is not affected by the presence of clouds, which are optically thin to the thermal emission. For 0.1 $\mu$m particles, the heating by ZnS is strong enough to evaporate KCl clouds in the dayside above 1 mbar (see Fig. \ref{figure_1}a and Fig. \ref{figure_2}). The evaporation of ZnS clouds stops this heating, producing a strong negative feedback which stabilizes the temperature below the ZnS evaporation curve (see Fig. \ref{figure_1}a). For 0.5 $\mu$m particles, the evaporation of KCl cloud is limited to pressures lower than 0.1 mbar.

\section{Observational spectra} \label{Observational spectra}

Using the GCM outputs, we produced transit, emission and reflection spectra and phase curves. The primary goal was to determine the planetary conditions required to match the observations of GJ1214b. The secondary goal was to reveal the best observational techniques and wavelengths to obtain information about the composition of GJ1214b's atmosphere.

\begin{figure*}[t] 
\begin{center} 
	\includegraphics[width=8.5cm]{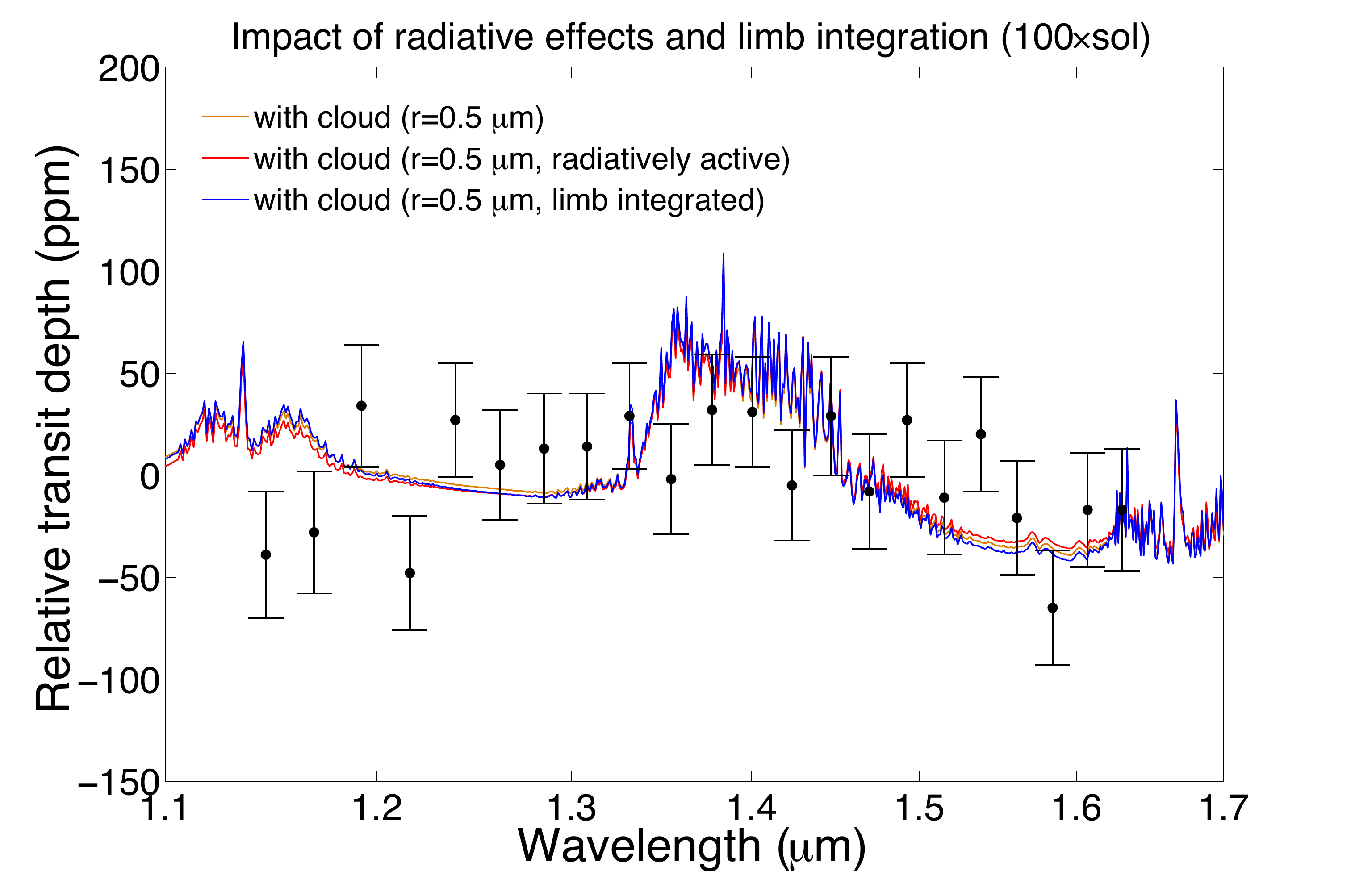}
	\includegraphics[width=8.5cm]{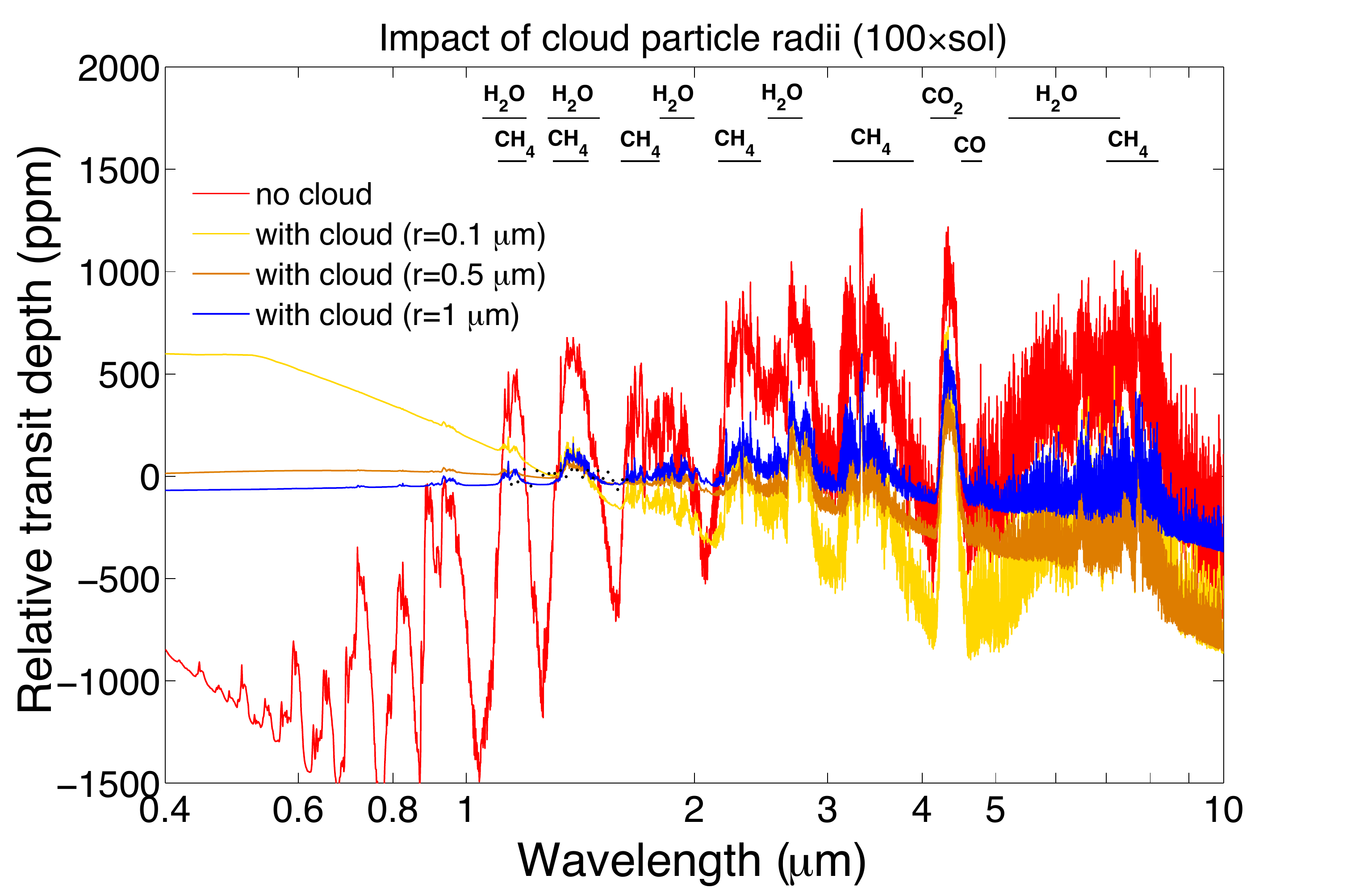}
	\includegraphics[width=8.5cm]{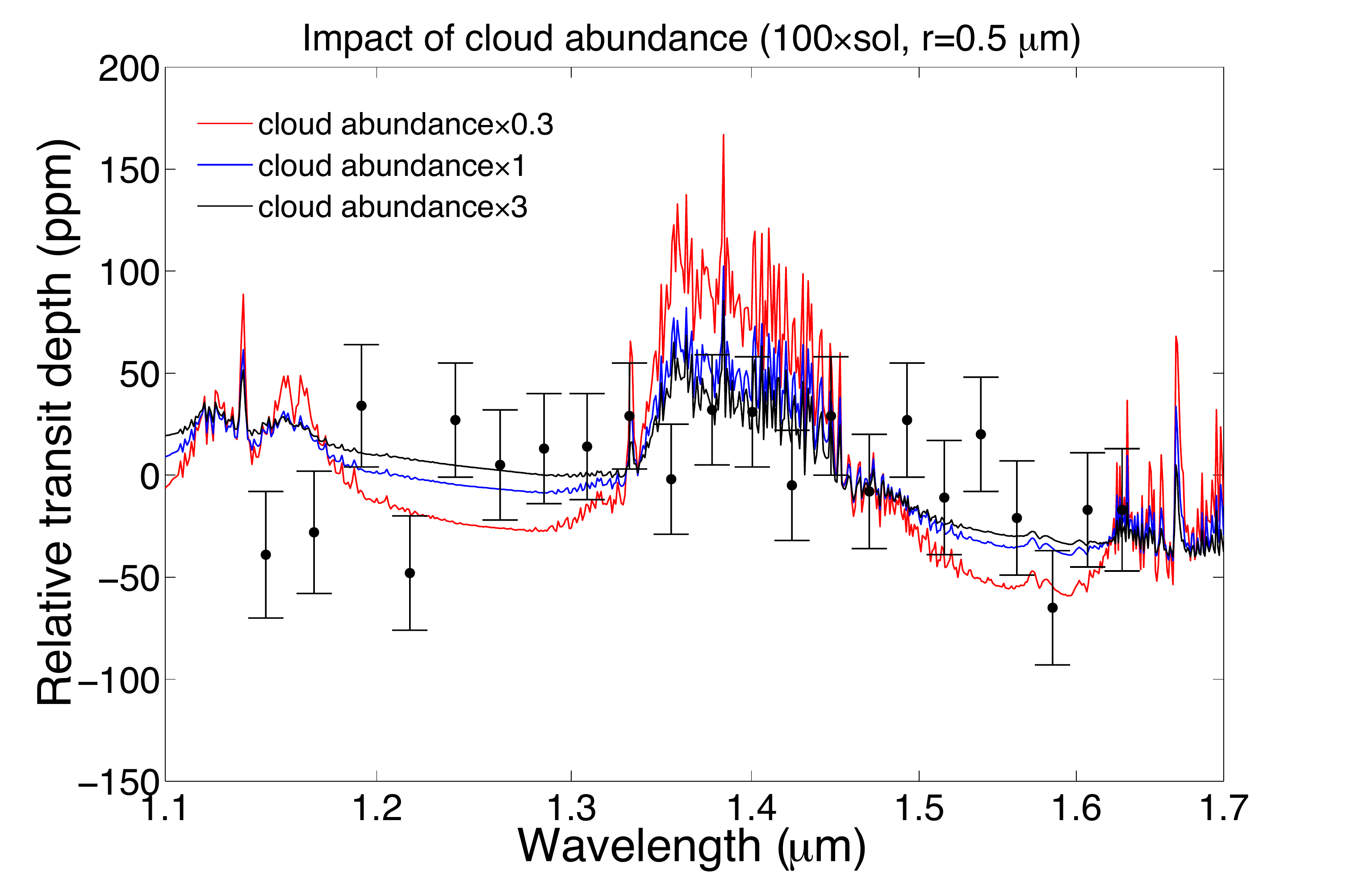}
	\includegraphics[width=8.5cm]{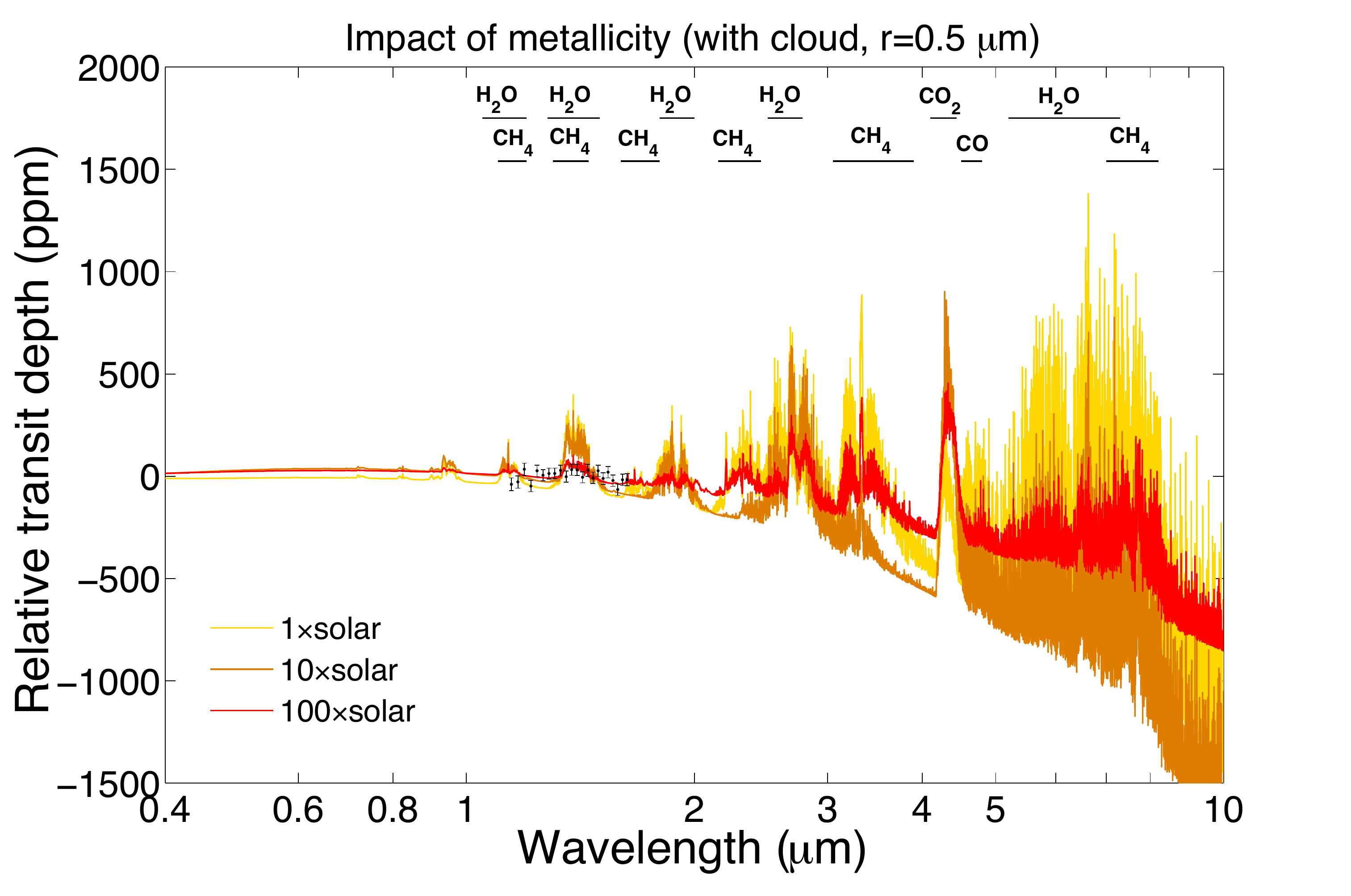}
\end{center}  
\caption{Transit spectra (relative transit depth in ppm) derived from the outputs of the GCM for non-radiatively active clouds. The top left panel shows spectra between 1.1 and 1.7$\mu$m with cloud particle radius of 0.5 $\mu$m, using the mean temperature and cloud profile at the limb (orange). 
The blue line is computed by integrating limb profiles at each GCM latitudinal point.
The red line is computed from the GCM simulation with radiatively active clouds.
The top right panel shows spectra between 0.4 and 10 $\mu$m without or with cloud (particle radii of 0.1, 0.5 and 1 $\mu$m). Major atmospheric molecular bands between 1 and 10 $\mu$m are indicated with black lines.
The bottom left panel shows spectra with cloud abundances multiplied everywhere by 0.3 (red) and 3 (black) compared to the reference case (blue).
The bottom right panel shows spectra for metallicity of 1, 10 and 100$\times$solar, with KCl and ZnS abundances scaled consequently.
In all panels, black dots correspond to data by \cite{kreidberg14a} with error bars.}
\label{figure_3}
\end{figure*}

\subsection{Transit spectra}
We computed transit spectra using the Spectral Mapping and Atmospheric Radiative Transfer code (SMART) \citep{meadows96, crisp97, misra14}. We computed spectra using HITRAN 2012 and using GCM temperature and cloud profiles, averaged at the limbs. Because the cloud inhomogeneities are weak at limbs, these averaged profiles are an excellent approximation to more accurate spectra obtained by combining apparent planetary radii from each latitudinal point at the limb (see Fig \ref{figure_3}a). 

Fig \ref{figure_3}a shows transit spectra with a cloud particle radius of 0.5 $\mu$m for the 100$\times$solar metallicity case. Differences between transit spectra from GCM simulations with radiatively and non-radiatively active clouds are negligible. The reference case (non-radiatively active cloud) always matches the data at less than 2$\sigma$ (reduced $\chi^2$=1.3 compared to 1.0 for a perfectly flat spectrum). In these simulations, the cloud-top pressure ($\tau_{transit}\approx1$) is around 0.02 mbar, 10-100 times higher than retrieved by \cite{kreidberg14a} for the same metallicity. Yet, there are several differences between these studies. Unlike our study, they use a gray cloud opacity, and their temperature and gravity are different (580 K and 8.48 m/s$^2$ versus our $\sim$460 K at limbs and 8.9 m/s$^2$). They also only consider water as a heavy element, so their mean molecular mass is smaller than ours (3.6 versus our 4.38 g/mol). These  changes make their atmosphere scale height 60$\%$ higher than ours, requiring higher clouds to get a flat spectrum.

Fig. \ref{figure_3}b shows transit spectra for the 100$\times$solar metallicity case without cloud, and with cloud particle radii of 0.1, 0.5 and 1 $\mu$m. Major atmospheric molecular bands between 1 and 10 $\mu$m are indicated in the figure. Particle radii of 0.3 $\mu$m or less lead to a strong slope in the visible and near-infrared and cannot match the observations at any abundance. For micrometric sizes (e.g. 0.5 and 1 $\mu$m), the transit spectrum is flat and featureless in the visible. In the infrared, large molecular features appear at wavelengths longer than around 3 $\mu$m (e.g. peak of CO$_2$ at 4.3 $\mu$m) where the clouds are optically thin. In addition, a slope can be clearly identified between 3 and 10 $\mu$m for 0.5 $\mu$m particles, whereas it remains very weak for 1 $\mu$m particles. According to our model, the cloud particle radius of 0.5 $\mu$m is the optimal size, allowing both strong mixing and a flatter transit spectrum with a reduced slope.

In Fig. \ref{figure_3}c, we tested deviations of cloud abundances compared to the reference case, assuming that it is proportional to KCl and ZnS vapor abundances in the deep atmosphere (as suggested by the limited impact of latent heat and cloud radiative effects on the vertical mixing). An enhancement by a factor 3 slightly flattens the spectrum. However, cloud abundances could be strongly limited by condensation occurring below 10 bars (see Fig \ref{figure_1}a).

Finally, Fig. \ref{figure_3}b shows transit spectra for the 1, 10 and 100$\times$solar metallicity cases, with 0.5 $\mu$m particles. Only the 100$\times$solar case can match the observational data.

\subsection{Emission spectra and phase curves}
\begin{figure*}[t] 
\begin{center} 
	\includegraphics[width=8.9cm]{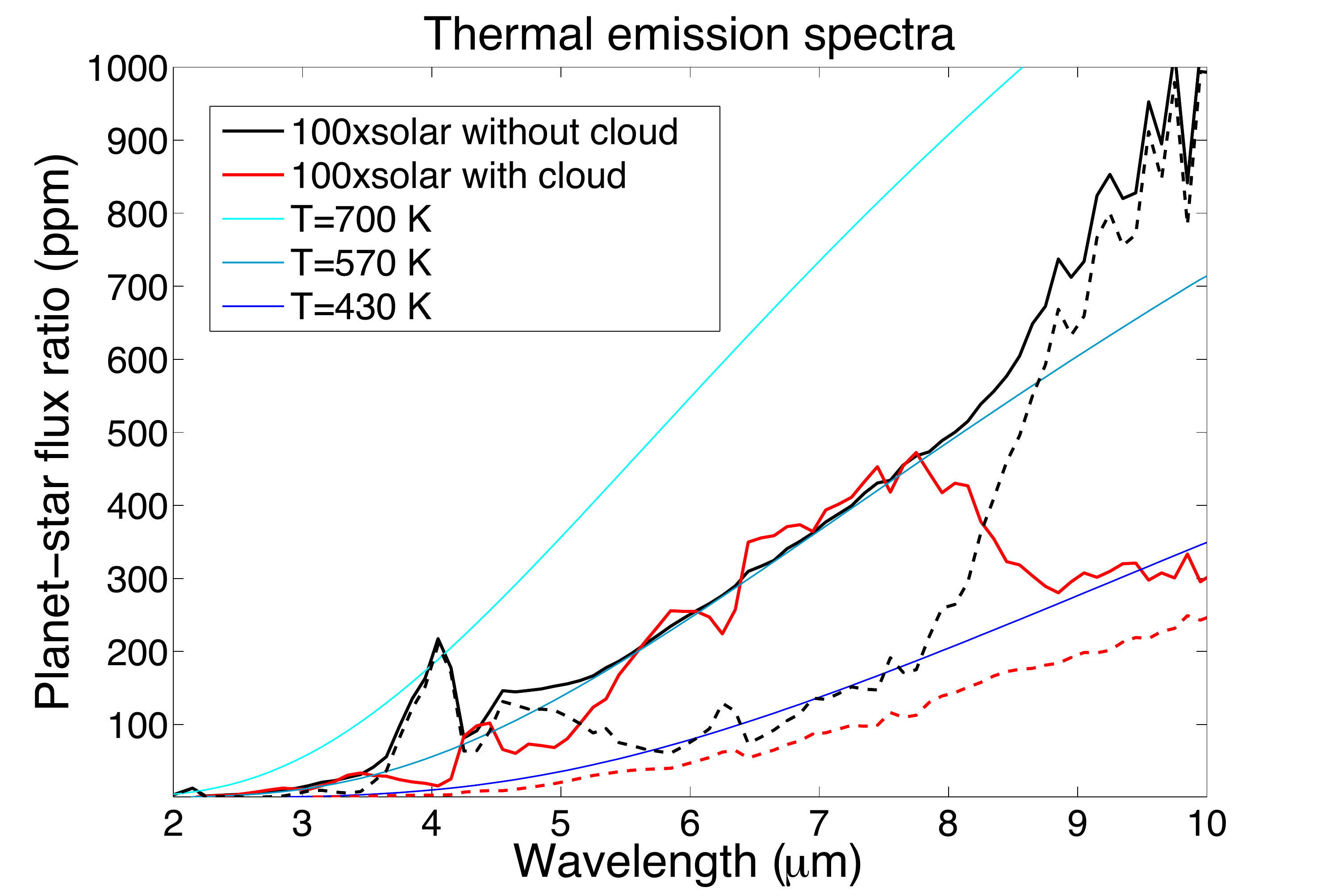}
	\includegraphics[width=8.8cm]{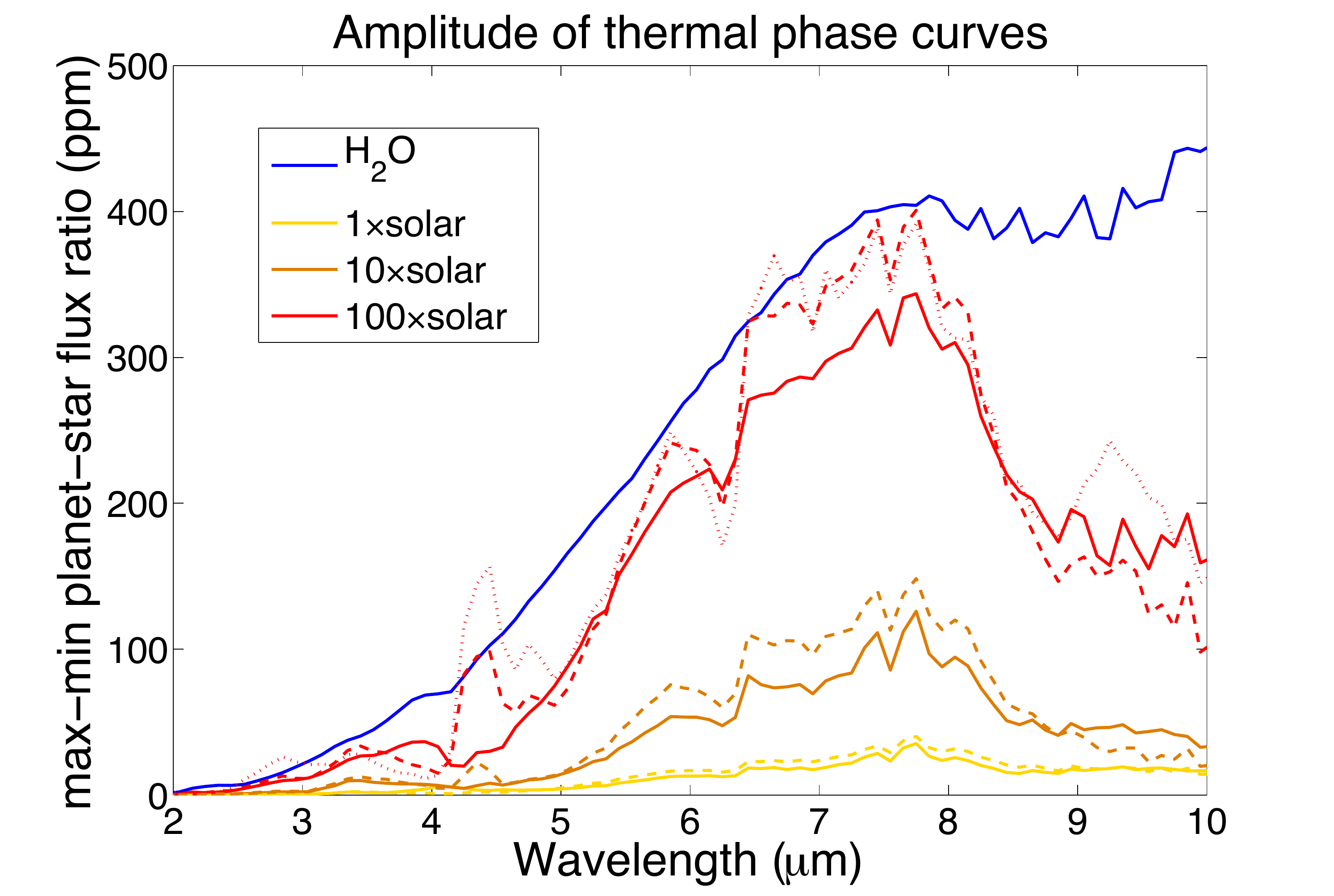}
\end{center}  
\caption{Thermal emission and phase curves with and without cloud.
Left panel shows the thermal emission with (red) and without (black) cloud with particle radius of 0.5 $\mu$m. Solid lines correspond to emission from the dayside and dashed lines to emission from the nightside. The blue lines are blackbody curves.
Right panel shows the amplitude of thermal phase curves without cloud (solid lines) and with clouds (dashed lines for radii of 0.5 $\mu$m and dotted line for radii of 0.1 $\mu$m) for metallicity of 1, 10, 100 and a pure water atmosphere.
For simplicity, we used a blackbody at 3026 K for the stellar flux.
}
\label{figure_4}
\end{figure*}

Fig. \ref{figure_4}a shows planetary emission spectra centered on the substellar (i.e. secondary eclipse) and the anti-stellar point (i.e. primary eclipse) for the 100$\times$solar metallicity, with and without cloud. These spectra were obtained by running the GCM for one day with a spectral resolution of 0.1 $\mu$m and using the GCM output fluxes directly. 
Without cloud, a strong emission peak with a brightness temperature of around 700 K is present at $\sim$4 $\mu$m corresponding to a water spectral window. A secondary peak is present at 4.5 $\mu$m and a third one at 8.2-10 $\mu$m corresponding to other spectral windows. The thermal emission strongly varies between the dayside and nightside in the water and methane absorption bands between 5.2 and 8.2 $\mu$m. At these wavelengths, the photosphere is located at around 0.5 mbar, where the temperature variations are strong (see  Fig. \ref{figure_1}a). 

0.5 $\mu$m cloud particles strongly impact the emission spectra. The thermal emission is generally reduced with clouds, except in the atmospheric absorption bands (e.g. 5.2-8.2 $\mu$m) where it is slightly enhanced. Clouds cool the atmosphere below 1 mbar and so the brightness temperature never exceeds 600K. The emission peaks at 4, 4.5 and 10 $\mu$m are reversed on the dayside because of the stratospheric thermal inversion. 
Globally, the presence of clouds has a limited impact on the amplitude of the phase curves (see Fig \ref{figure_4}b), which primarily depends on the atmospheric metallicity \citep{menou12}. 
Therefore, the observation of thermal emission variations, in particular in the 6.3 $\mu$m water band (between 5.2 and 7.3 $\mu$m), is an excellent probe of the atmospheric metallicity of GJ1214b.

\subsection{Reflection spectra}
\begin{figure}[!h] 
\begin{center} 
	\includegraphics[width=8.9cm]{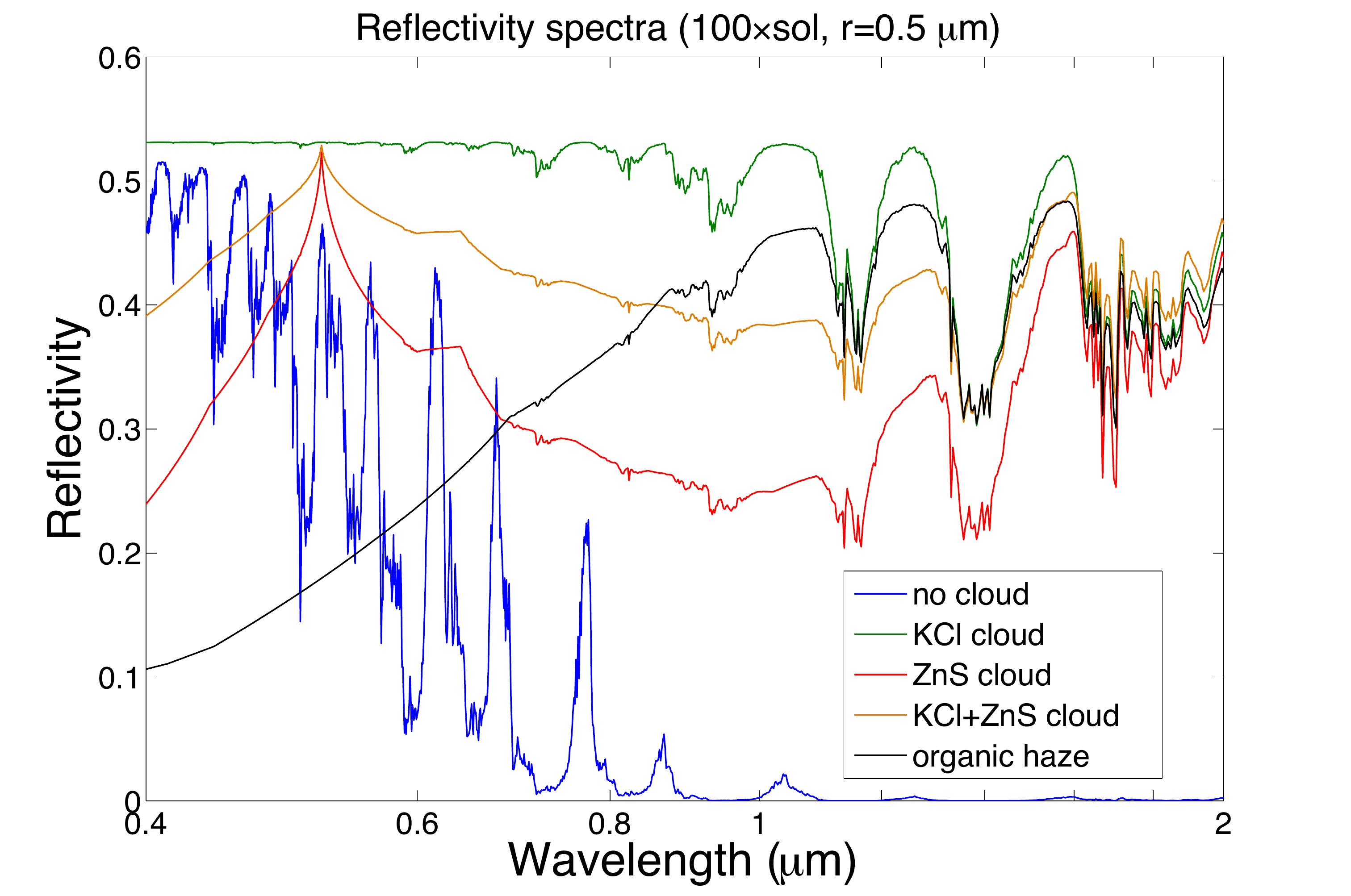}\\
	\includegraphics[width=8.2cm]{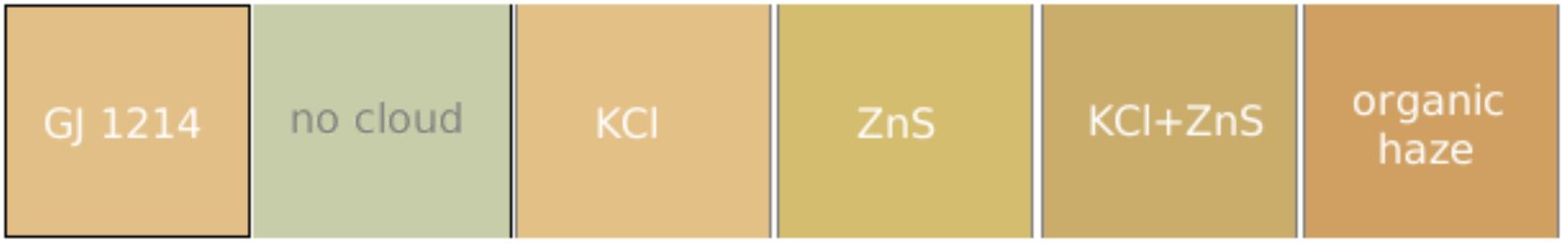}
	\includegraphics[width=8.2cm]{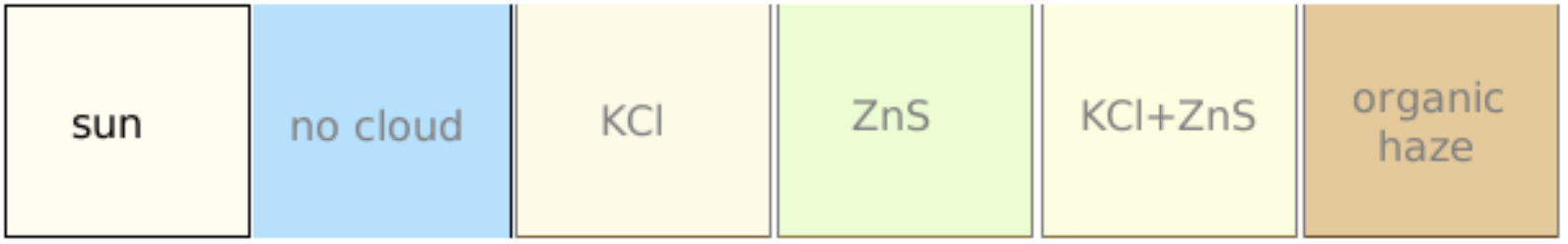}
\end{center}  
\caption{Top panel: reflectivity spectra with clouds (KCl only, ZnS only, KCl+ZnS or organic haze) or without cloud using GCM results for 0.5 $\mu$m particles and assuming a stellar zenith angle of 60$^{\circ}$. Organic haze particles were assumed spherical with the same concentration as KCl clouds.
Bottom panel: color of GJ1214b orbiting a GJ1214-like M dwarf (AD Leo, top), and orbiting a Sun-like star (bottom) calculated by convolving spectral brightness with eye response (www.brucelindbloom.com).}
\label{figure_5}
\end{figure}

Fig. \ref{figure_5}a shows reflection spectra for a non-cloudy atmosphere with 100$\times$solar metallicity and for cloudy atmospheres with KCl, ZnS or Titan-like organic haze particles. We used GCM profiles for 0.5 $\mu$m particles. 
KCl clouds are purely scattering and produce a constant reflectivity in the visible. ZnS clouds absorb over most of the visible but do not absorb at 0.5 $\mu$m, producing a peak of reflectivity.
Organic haze strongly absorbs at short visible wavelengths. For wavelengths higher than around 1 $\mu$m, strong atmospheric molecular bands of water and methane are present.

Using these reflection spectra, we evaluated the apparent (human eye) colors of GJ1214b for clouds or haze illuminated by a GJ1214-like star and the Sun (see Fig. \ref{figure_5}b). If cloudy, GJ1214b would appear orange. Orbiting around a Sun-like star, a cloudy GJ1214b could appear white, green or orange.

\section{Discussion} \label{Discussion} 

According to our model, micrometric cloud particles could be lofted to the upper atmosphere of GJ1214b with a very strong impact on transit spectra. However, a transit spectrum consistent with HST observations could only be obtained for cloud particles with a radius around 0.5 $\mu$m. In reality, clouds should have a size distribution depending on altitude and latitude. Larger particles would sediment faster and decrease the amount of cloud in the upper atmosphere, whereas smaller particles would produce a stronger slope in the transit spectrum. A more realistic cloud distribution would likely produce a less flat transit spectrum. However, other effects such as metallicity and photochemistry may help to flatten the spectrum. 
A higher metallicity than the 100$\times$solar considered here would require a lower cloud-top \citep{kreidberg14a} and imply a higher amount of cloud but still with a similar vertical mixing. It could efficiently flatten the transit spectrum.
Also, we assumed thermochemical equilibrium in this study. However, the absorption bands seen between 1 and 2 $\mu$m in the cloudy transit spectra (Fig \ref{figure_3}) come predominantly from pressures lower than 0.01 mbar. At such low pressures, the atmosphere is far from chemical equilibrium, and water and methane can be strongly photolyzed \citep{miller-ricci12}. 
While other absorbing species are produced or enhanced, the decrease of the two major absorbers may significantly flatten the transit spectrum between 1 and 2 $\mu$m.

If we assume that JWST will reach 30$\%$ of the photo noise limit (see \cite{koll15}), the 3$\sigma$ uncertainty for one primary or secondary transit (duration around 50 min) will be around 95 ppm for the CO$_2$ band at 4.2-4.4 $\mu$m, around 120 ppm for the H$_2$O band at 5.2-7.3 $\mu$m and around 185/170 ppm for the window wavelength ranges 4.0-4.1/9-10 $\mu$m.
According to Fig. \ref{figure_3}, one transit of GJ1214b observed by JWST should therefore be enough to detect molecules (i.e. H$_2$O, CO$_2$ and CH$_4$) and the Mie slope for particle radii lower than 1 $\mu$m. 
According to Fig. \ref{figure_4}, one secondary eclipse should allow detection of the thermal inversion (both at 4 and 10 $\mu$m), while one full phase curve should provide a good estimation of the atmospheric metallicity. The combination of phase curve and secondary eclipse would allow estimation of the bond albedo \citep{cowan11a}.
The observation of a few primary/secondary eclipses or full orbits by JWST could provide very precise spectra and phase curves revealing GJ1214b's atmospheric composition and providing clues on the size and optical properties (i.e. absorbing or not) of clouds. Non-absorbing clouds would suggest KCl particles. Absorbing clouds would favor ZnS particles or organic haze. In that case, the best way for determining the composition of cloud particles would be direct imaging or secondary eclipses/phase curves of reflected light in the visible. The different clouds/haze have characteristic features in visible reflectivity spectra. Future large telescopes such as ELT may have the capabilities for measuring this.

Finally, our results concerning cloud distribution, radiative effects and impacts on spectra for GJ1214b are very general and can be applied to many warm mini-Neptunes. Some current missions (e.g. K2) and future missions (e.g. TESS and PLATO) should detect several GJ1214b-like planets. A global survey of these objects by future telescopes (e.g. JWST, ELT or future EChO-class telescopes) would provide insight into the nature and origin of clouds on extrasolar planets by collecting statistical information on the atmospheric conditions required for cloud and haze formation.

\acknowledgments
\paragraph{Acknowledgments:} 
We are grateful to Eddie Schwieterman for help concerning the use of SMART and to Dorian Abbot for a helpful review.
B.C. acknowledges support from an appointment to the NASA Postdoctoral Program,   administered by Oak Ridge Affiliated Universities.
This work was performed as part of the NASA Astrobiology Institute's Virtual Planetary Laboratory, supported by NASA under Cooperative Agreement No. NNA13AA93A.
This work was facilitated though the use of the Hyak supercomputer system at the University of Washington.


\bibliographystyle{apj}

\end{document}